\documentclass{aa}
\usepackage{graphicx}
\newcommand{\Ha}{H$_{\alpha}{ }$}
\newcommand{\Hb}{H$_{\beta} $ }
\newcommand{\Hg}{H$_{\gamma}{ }$ }
\newcommand{\Hd}{H$_{\delta}{ }$}
\newcommand{\HeII} {HeII(~4686\,\AA) }

\newcommand{\rx}{V709~Cas }

\begin{document}                                          
\title{The white dwarf revealed in the intermediate polar V709 Cassiopeiae.
\thanks{Based on data collected at the Haute-Provence Observatory OHP
(France)}}
\author{J.M. Bonnet-Bidaud\inst{1} 
\and M. Mouchet\inst{2,3}
\and D. de Martino\inst{4} 
\and G. Matt\inst{5}
\and C. Motch\inst{6}}
\offprints{J.M. Bonnet-Bidaud}
%\mail{bobi@discovery.saclay.cea.fr}
\institute
{Service d'Astrophysique, DSM/DAPNIA/SAp, CE Saclay, 
F-91191 Gif sur Yvette Cedex, France\\
\email{bobi@discovery.saclay.cea.fr}
\and
DAEC et UMR 8631 du CNRS, Observatoire de Paris Section de Meudon, F-92195 Meudon Cedex, France
\and
Universit\'{e} Denis Diderot, Place Jussieu, F-75005 Paris, France
\email{martine.mouchet@obspm.fr}
\and
Osservatorio Astronomico di Capodimonte, I-80131 Napoli, Italy
\email{demartino@na.astro.it}
\and
Dipartimento di Fisica, Universita Roma Tre, I-00146 Roma, Italy
\email{matt@fis.uniroma3.it}
\and
CNRS,  Observatoire de Strasbourg, 
F-67000 Strasbourg, France
\email{motch@astro.u-strasbg.fr}
}
\date{Received date: 3 April 2001 ; accepted date: 23 May 2001}
\titlerunning{The white dwarf in V709 Cas}
%\authorrunning{Bonnet-Bidaud et al.}

\abstract{
Results are presented from the first detailed spectroscopic observations of 
the recently identified intermediate polar RXJ0028.8+5917/V709 Cas. 
The study of the emission line radial velocities allows us to remove the uncertainties
on the different aliases of the orbital period and a best value is found 
at (0.2225$\pm$0.0002)\,day.
It is also found that the system shows significant EW$\sim$ (2-4)\,\AA\, broad 
absorptions affecting the Balmer lines from H$_{\delta}$ to H$_{\beta}$. 
These broad absorptions are interpreted as the contribution of an underlying white 
dwarf atmosphere.
The characteristics of the absorptions are found to be consistent with a 
DA log~$g$\,=\,8 white dwarf at a temperature of $\sim$23\,000\,K, contributing 
 $\sim$17\% (at 4500\,\AA) to the overall flux.
This is the first direct detection of a white dwarf in an intermediate polar system.
The absence of significant Zeeman splitting indicates a magnetic field lower 
than 10 MG, confirming that, at least in some cases, intermediate polars have 
weaker fields than polars. Different possibilities are discussed to explain
the substantial contribution of the white dwarf to the overall flux.
\keywords{stars: white dwarfs -- stars: cataclysmic variables -- stars: magnetic fields
-- x-rays: individuals: RXJ0028.8+5917}
 } 

\maketitle

\section{Introduction}
Cataclysmic Variables (CVs) consist of a white dwarf 
accreting matter from a low mass companion filling its Roche lobe
which, in some cases, can be an evolved secondary.
Among them, polars and intermediate polars, or magnetic CVs,
are a sub-class where the magnetic field of the white dwarf strongly affects
the mass flow within the binary system (see reviews by Cropper 1990
and Patterson 1994).
In polars (also called AM Her stars), the field is strong enough
to lock the white dwarf spin in synchrony with the orbit
and to control the flow into an accretion column.
In Intermediate Polars (IP, previously DQ Her stars), the field is thought to be 
weaker, leading to a desynchronisation with a white dwarf spin faster than the 
orbit. The accretion may then take place either through an accretion disc (disk-fed),
or directly from the inner Lagrangian point (diskless) or via a combination 
of both (overflow-disk) (see review by Patterson 1994).
Since CVs are interacting binary systems, their emission from the X-rays to the infrared
is generally totally dominated by the 
release of the gravitational energy of the accreted matter in a disk and/or an accretion 
column and not by the emission of the stellar components (white dwarf and low-mass companion).
In polars, some hints of the stellar characteristics can however be obtained during
the period of low accretion rates which is a typical feature of these systems.
During these low states, detected stellar photospheric lines can be used to constrain
the temperature and magnetic field of the white dwarf (see Beuermann 1996, Sion 1999) 
as well as the nature and distance of the companion (see Cropper 1990, Warner 1995a).
Possibly due to a different evolutionary stage and to a higher accretion rate,  among IPs no 
such low states are known to exist and therefore very little is known about the
stellar characteristics. The only cases where a stellar spectrum has been detected are
for the very peculiar long period AE Aqr and GK Per systems where the secondary is an evolved 
star (Welsh et al. 1993, Reinsch 1994).

The X-ray source RXJ0028.8+5917 was detected in the ROSAT All Sky Survey and
identified as an intermediate polar on the basis of its spectral characteristics
with a hard X-ray ($\sim$10 keV) component and a N$_{H}$ $\sim 10^{21}$ at.cm$^{-2}$ column density 
(Haberl \& Motch 1995). Further pointed observations by the ROSAT satellite led to the 
detection of an X-ray pulsation at 312.8\,s with a 40$\%$ amplitude and a probable 
identification with coincident catalogued sources from previous different surveys 4U0027+59, 3A0026+593 
and 1H0025+588 (Haberl \& Motch 1995).
Subsequent optical observations identified the source with the m${_v}$~$\sim$~14 
variable star, V709~Cas, and line velocity variations reveal two possible orbital
periods P\,=\,5.4$\pm$0.2\,h and P\,=\,4.45$\pm$0.15\,h (Motch et al. 1996, hereafter M96).
Optical oscillations at the X-ray spin period and sideband have also been reported 
(Kozhevnikov 2001).
A ROSAT observation in February 1998 showed a peculiar 
double-peaked X-ray pulsation with peaks separated by one third of the spin period
and the source is suspected to be fed by accretion through a disk seen at low
inclination angle and with a relatively weakly magnetized white dwarf 
(Norton et al. 1999). A more detailed spectral analysis has been provided by 
recent BeppoSAX and RXTE observations (de Martino et al. 2001a, submitted)

We report here on spectroscopic observations of the intermediate polar 
RXJ0028.8+5917/V709 Cas, which clearly shows the evidence of broad Balmer absorption lines,
superimposed on the usual IP emission line spectrum. These lines are attributed 
to the white dwarf atmosphere and are used to constrain for the first time
the temperature and magnetic field of the white dwarf in an intermediate polar
system.

\section{Observations and Reduction}
Spectroscopic observations of V709 Cas were  carried out at the 
1.93m telescope of the Haute-Provence Observatory (OHP),
equipped with the Carelec spectrograph (Lemaitre et al. 1990).
A 130\,\AA/mm grating was used to obtain low resolution (FWHM $\approx$ 7\,\AA) 
spectroscopy over a wide (3600 - 7200\,\AA) range.
Observations of (2-3) hour duration were performed in three consecutive nights and 
an additional one separated by one day (see Table 1).
To avoid interruption in the orbital coverage, wavelength and photometric absolute
calibrations were performed at the beginning and end of the run. A careful check for the
stability of the wavelength calibration during the run was made, using the strongest
sky emission lines present in the spectrum, such as the O [I] line (\,5577.34\,\AA),
and small variations were corrected to an accuracy better than 0.1\,\AA.
The standard stars BD +28$^{\circ}$4211 and Feige 110 were used for the photometric calibration 
on each night but due to variable atmospheric absorption, only the night of Aug. 23 could 
be used to provide reliable absolute fluxes.
To smooth out the influence of a possible modulation with the short ($\sim$ 5\,min) 
spin period, spectra were accumulated in either 15\,min or 20\,min exposures.
All spectra were corrected for bias, flat-fielded and wavelength calibrated using 
the standard ESO/MIDAS procedures.

\begin{table}
\caption[ ]{Log of observations}
\begin{flushleft}
\begin{tabular}{lrrrr}
\hline
\multicolumn{1}{c}{Date} & \multicolumn{1}{c}{Tstart(*)} & 
\multicolumn{1}{c}{Tend(*)} & \multicolumn{1}{c}{n} & \multicolumn{1}{c}{T(min)} \\
\hline
 & & & \\
1998 Aug. 21 & 1047.47164 & 1047.64034 & 9  & 20 \\
1998 Aug. 22 & 1048.42757 & 1048.64590 & 11 & 20 \\
1998 Aug. 23 & 1049.49154 & 1049.64089 & 8  & 20 \\
1998 Aug. 25 & 1051.45438 & 1051.64903 & 15 & 15 \\
\hline
\end{tabular}
\end{flushleft}
(*) HJD(2450000+), n number of spectra, T exposure times
\end{table}

\section{Analysis and results}
\subsection{The two-component spectrum}
A total of 43 spectra were acquired over the 4 nights of observations.
The spectra on the different nights are similar to those reported
from the first optical observations by M96, showing strong
Balmer and \HeII  lines in emission, superimposed on a blue continuum.
%Figure~\ref{fig1sp} 
Figure 1 shows a representative mean spectrum, averaged over one night.
Weaker HeI and HeII emission lines are also visible in the figure.
The bump in the continuum, visible around 5500\,\AA\,, is due to broad variable atmospheric
extinction due to ozone, not corrected by the applied standard extinction curve.
Absorptions around 6280\,\AA\, and 6880\,\AA\, are also atmospheric features.
Absorption on the red side of HeI (\,5875\AA) is most likely due to the NaD
doublet. Its origin may be interstellar but it may also reveal a significant
contribution from the secondary.

\begin{figure*}[ht]
%\sidecaption
%\includegraphics[angle=-90,width=12cm]{rx28fig1.ps}
%\resizebox{12cm}{!}{\includegraphics[angle=-90]{rx28fig1test.ps}}
\resizebox{12cm}{!}{\includegraphics{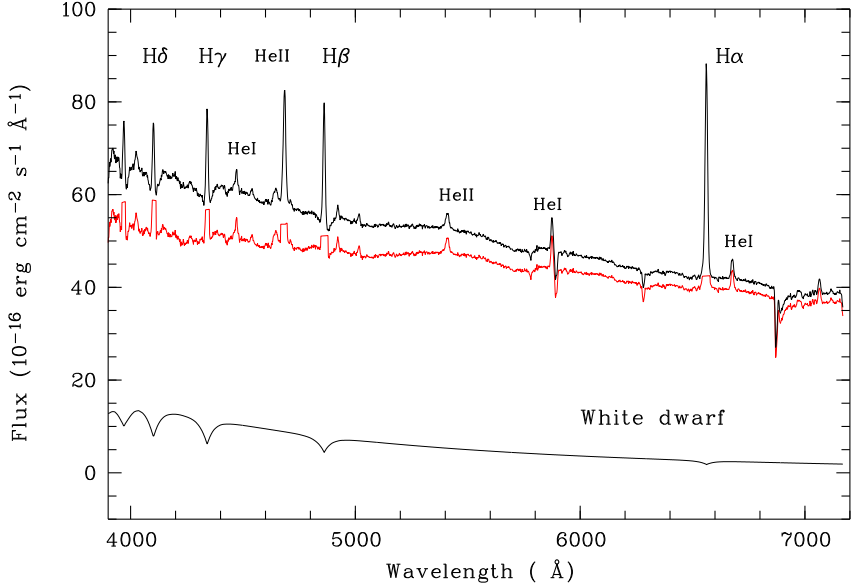}}
\hfill
\parbox[b]{55mm}{ \caption{The mean spectrum of V709 Cas on Aug. 23. The top curve is 
the observed spectrum
showing a significant wide absorption affecting the Balmer lines from \Hd\, to \Hb.
The middle curve is the residual flux after subtracting a contribution from
a T=23\,000K white dwarf whose flux distribution is shown by the bottom curve.
Emission lines in the middle curve have been cut for clarity. The spectra 
have not been corrected for the telluric absorption lines. The narrow absorptions and
the broad bump around 5500\,\AA\, are atmospheric. The narrow absorption feature in
the red side of HeI (5875\,\AA) is identified as NaD (see text).
}
\vspace{1cm}}
\label{fig1sp}
\end{figure*}

The overall spectrum of V709 Cas is typical of IPs but a more careful inspection 
of the lines reveals an unusual feature. 
Beside the presence of an emission component, the Balmer lines show a very
significant symmetrical depression on both sides of the emission, suggestive of a 
broad absorption. This feature is absent in the \HeII line and the \Ha\, line. 
The spectra obtained during the different nights all show the same characteristics, 
though it is slightly more pronounced on Aug. 23.  
The normalized source spectrum is presented in Fig. 2,
%Fig.~\ref{fig2sp} 
which shows more clearly that the most significant absorption is seen at \Hg and \Hd.
The feature is less symmetric around \Hb where the 
blue part of the absorption seems to be  filled up.

\subsection{The orbital period}
To take into account the absorption features, the line characteristics 
(velocities, equivalent widths), have been derived 
by a fit with a two-gaussian 
function for emission and absorption,
with the adjacent continuum being described by a one degree polynomial. 
The velocities of the strongest \Ha\, emission line, measured over the
5 day interval, have been used to determine the orbital period. 
A Fourier transform (FFT) of the velocities was first used to constrain the period.
The FFT yields two prominent peaks at periods
P\,=\,5.34$\pm$0.05\,h and P\,=\,4.37$\pm$0.06\,h with smaller peaks at 
P\,=\,3.70$\pm$0.06\,h and P\,=\,6.87$\pm$0.06\,h, all one day
aliases with respect to each other, the errors being determined by the width of the
peak at half maximum. The two most significant periods
are in accordance with the values reported by M96.

Velocities of the emission lines were then folded with the two 
most significant periods and fitted with a function 
V~=~$\gamma$\,+\,K.sin[(2$\pi$t/P)-$\phi_{0}$]
with $\gamma$, K, and $\phi_{0}$ as free parameters, 
using a $\chi^{2}$ minimization program. For all lines, the 
fit is significantly better at P\,=\,5.34\,h, so that the main alias period
of P\,=\,4.37\,h can be securely rejected. For H$_{\alpha}$, the reduced $\chi^{2}_{r}$ value decreases
from 9.8 (P\,=\,4.37\,h) to 4.0 (P\,=\,5.34\,h), corresponding 
to a significance at a level better than 99\%, computed from a F-test.
The (O-C) values of the sine blue-to-red crossing times, computed 
with the trial period of 5.34\,h period, were then fitted
to yield a best period value 
of P~=~(5.341$\pm$0.005) h with a corresponding ephemeris as
T$_o$~=~HJD2451048.0575(2) + E*0.2225(2)\,d,
where T$_o$ is the blue-to-red crossing time of the emission lines.

\begin{figure}
\resizebox{\hsize}{!}{\includegraphics[angle=-90]{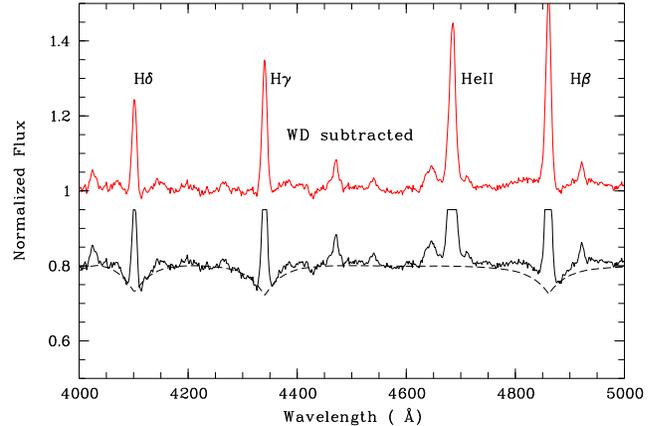}}
\caption{
A close-up of the average normalized spectrum of \rx showing clearly the 
absorptions around \Hd, \Hg and \Hb 
(lower curve, shifted by -0.2 unit for clarity). 
Note the absence of such a feature
around \HeII. 
The dotted line shows the contribution of the best fit white dwarf (see text).
The upper curve is the residual after subtraction of the white dwarf contribution.
}
\label{fig2sp}
\end{figure}

\subsection{The line characteristics} 

\begin{table}
\caption[ ]{Emission line radial velocities}
\label{vel}
\begin{flushleft}
\begin{tabular}{lrrrr}
\hline
Line & \multicolumn{1}{c}{$\gamma$(km.s$^{-1}$)} & \multicolumn{1}{c}{K(km.s$^{-1}$)}  
& \multicolumn{1}{c}{$\phi_{0}$} & \multicolumn{1}{c}{$\chi^{2}_{r}$} \\
\hline
 & & & &  \\
H$_\alpha$ & $-$64.7 (1.5) & 102.6 (2.3) & 0.000(3) & 4.0  \\
H$_\beta$  & $-$56.7 (1.5) & 92.7 (2.3)  & -0.022(4) & 4.0  \\
H$_\gamma$ & $-$37.6 (1.5) & 88.9 (2.3)  & -0.024(4) & 5.8  \\
H$_\delta$ & $-$51.0 (1.5) & 77.1 (2.3)  & -0.008(4) &12.8  \\
HeII(4686\AA) & $-$91.1 (1.5) & 75.5 (2.3)  & +0.044(4) & 7.2  \\
\hline
\multicolumn{5}{l}{Errors on last digit in parentheses.}
\end{tabular}
\end{flushleft}
\end{table}

%Table~\ref{vel} 
Table 2 
gives the amplitude (K) and 
phasing ($\phi_{0}$) of the different emission lines
folded with the above ephemeris.
All Balmer lines show velocity curves with similar phasing but 
with radial velocity amplitudes decreasing through the Balmer series from 103 km.s$^{-1}$ 
(\Ha) to 77 km.s$^{-1}$ (\Hd).
The \HeII shows the lowest amplitude at $\sim$~76 km.s$^{-1}$ 
with a slight shift in phase and a different $\gamma$ velocity.

\begin{figure*}[ht]
%\resizebox{12cm}{!}{\includegraphics[angle=-90]{rx28fig3.ps}}
\resizebox{12cm}{!}{\includegraphics{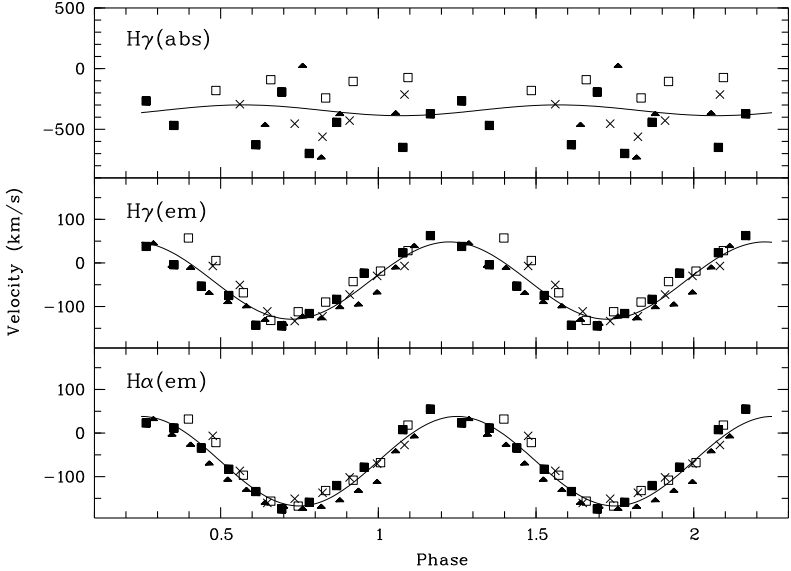}}
\hfill
\parbox[b]{55mm}{
\caption{
Radial velocity curves as a function of orbital phase
for the \Ha (bottom) and \Hg (middle) emission components and
the \Hg (top) absorption component. The different nights are shown
by different symbols 
(Aug 21-squares, Aug 22-filled squares, Aug 23-crosses, Aug 25-filled triangles)
and the best sinusoidal fits are shown by thin lines.
Typical errors are (3-5) km.s$^{-1}$ (bottom), (5-10) km.s$^{-1}$ (middle)
and (100-200) km.s$^{-1}$ (top).}
\vspace{1cm}
}
\label{fig3vel}
\end{figure*}

The velocity curves for H$\alpha$ and H$\gamma$ emission lines are shown 
%in Figure~\ref{fig3vel}. 
in Figure~3. 
Inspection of the curves reveals a significant dispersion of
the points obtained on the different nights. Though the minimum around $\phi$
$\sim$\,0.75 is relatively well defined, there is a noticeable shift 
in velocity
from one night to another 
for points at the same phase, in the range $\phi$ $\sim$(0.3-0.5) and (0.8-1.0).
The dispersion from a pure sine curve is also reflected by the high values 
of the reduced $\chi^{2}_{r}$ of the fits (Table 2). 
Using the sky emission lines, we checked that between the different nights the
accuracy of the absolute calibration was better than 0.1\,\AA.
The shift in velocity from night to night is of the order of 70 km.s$^{-1}$ 
(1.0\,\AA{ } at \Hg and 1.5\,\AA{ } at \Ha).
There is no clear change in the line profiles from night to night but we 
cannot exclude the existence of some asymmetry in the lines that would be 
poorly fitted by gaussian curves. 
Fits of the emission lines with an unique gaussian function 
yield a similar amount of dispersion, showing that the effect is not due
to the existence of the absorption component. The distortion is also present in
\Ha\, where no absorption is seen.
The influence of the spin period can also be 
excluded as the exposure times of 20 min and 15 min were chosen to be
close multiples of the spin period (3.8 and 2.9 respectively).
Possible significances of this dispersion are considered in the discussion. 

The velocity of the \Hg absorption component, as deduced from the fits, is also
plotted in Figure 3. The measured values are however highly uncertain due to the
very shallow and broad nature of the line and no significant radial velocity 
curve can be derived. An attempt was made to measure the velocity changes
with a cross-correlation method, using one of the best S/N spectra as a template.
Though somewhat reducing the dispersion of the points, this method also does not 
reveal any regular variations. 

\begin{table}
\caption[ ]{Line full widths at half maximum (FW) and equivalent widths (EW)}
\label{EWFW}
\begin{flushleft}
\begin{tabular}{lrrrr}
\hline
Line & \multicolumn{2}{c}{Emission} & \multicolumn{2}{c}{Absorption} \\
 & EW$_{e}$(\AA) & FW$_{e}$(\AA) &  EW$_{a}$(\AA) & FW$_{a}$(\AA) \\ 
\hline\\
H$_\alpha$ & 18.0(4) & 14 (1)     & -          &  -     \\
H$_\beta$  & 6.6(3)  & 12 (1)     & 1.5 (0.2)  & 36 (5) \\
H$_\gamma$ & 4.7(3)  & 11 (1)     & 3.5 (0.1)  & 58 (7) \\
H$_\delta$ & 3.8(3)  & 11 (1)     & 3.0 (0.1)  & 37 (5) \\
HeII(4686\AA) & 6.7(3)  & 13 (1)     &  -         & -      \\
\hline
\multicolumn{5}{l}{Errors on last digit in parentheses.}
\end{tabular}
\end{flushleft}
\end{table}

%Table~\ref{EWFW} 
Table 3 gives the measured equivalent widths (EW) and intrinsic widths (FW at half-maximum) 
of the emission and absorption components.  
For the Balmer lines where an absorption is measured, 
the absorption is always very broad (FWHM $\sim$ 2500-4000 km.s$^{-1}$)
and with an EW nearly comparable to emission, with the exception of \Hb
for which the EW is significantly smaller, due to the partial filling of the
blue part of the absorption. 
The intrinsic widths of the absorption component ($\sim$40-60\,\AA)
are similar to what is observed in a spectrum of a DA white dwarf, 
with Balmer lines in absorption. 
 
\section{Discussion}
The optical spectrum of \rx, with strong (EW$\sim$4-18\,\AA)
Balmer and \HeII emission lines, is typical of IPs. However, it also
shows evidence of broad (FWHM$\sim$40-60\,\AA) absorptions which were not
yet noticed in this source nor in any other IP.
\,Balmer lines in absorption are sometimes seen among classical CVs, 
mainly in nova-like systems in high states
and during dwarf nova eruptions, which suggests that they are
formed in an optically thick accretion disk with a high mass transfer
(see for instance La Dous 1994, Hessman 1986). 
Absorption lines have also been observed among dwarf novae in a very few systems where
the accretion rate is low, as in GW Lib (Szkody et al. 2000) or with a high inclination,
as in OY Car (Hessman et al. 1989).
The other known cases are among polars during
low states where Balmer absorption lines are clearly seen, significantly split
by the Zeeman effect. They are attributed in this case to the atmosphere of 
a strongly magnetized white dwarf (see Cropper 1990).
In the case of V709 Cas, the stability of the optical flux excludes a dwarf
nova event and points toward a white dwarf rather than a disk origin. 
The absence of He absorptions also does not favour a disk origin. 
The lack  of any significant absorption around \Ha\, is indirect
evidence that the spectral component responsible for the absorption 
has a rather steep continuum, contributing significantly only in the blue
and becoming negligible in the red part of the spectrum by a dilution effect 
with the rest of the emission. This is therefore indicative of a hot white 
dwarf as commonly seen among polars.\\
We therefore attempted to describe the underlying absorption spectrum, 
using a grid of WD
models with LTE atmospheres (Finley et al. 1997, Koester 2000, private communication). 
As the contribution of the white dwarf to the overall flux is unknown, 
we used an indirect method.
The temperature-gravity domain was first constrained by comparing the observed 
intrinsic line widths with the ones predicted from theoretical models. The
contribution of the white dwarf to the overall flux was then deduced from
the measured EW of the absorption lines. 

For the determination of the observed parameters, we use primarily 
the \Hg measurements as the \Hd\, line is weaker and the \Hb line seems
to be contaminated by side emissions.
%Figure~\ref{fig4rel} 
Figure~4 shows the theoretical FWHM for a grid of LTE models 
of DA white dwarfs with a temperature 
ranging from T = 15\,000 to 45\,000\,K and a gravity from log~$g$~=~7.5 to 8.5.
Assuming a gravity of log~$g$~=~8, the measured FWHM given in Table 3
%Table~\ref{EWFW}
is consistent with a temperature in the range T~=~18\,000-30\,000\,K with a best value
at 23\,000\,K. If one allows for a full range of gravities, the T-log~$g$ domain is
not unambiguously determined and the temperature is only weakly constrained
in the range 15\,000-35\,000\,K.
 
A dilution factor was defined as the ratio of the white dwarf 
emission to the global continuum flux, 
$\alpha$= F$_{wd}$/F$_{cont}$. For a continuum flux close to an emission line,
$\alpha$ is also the ratio EW$_{abs}$/EW$_{th}$, where  EW$_{abs}$ is the observed 
EW of the absorption (including dilution) and EW$_{th}$ the theoretical EW value. 
The dilution factor and hence the contribution of the white dwarf to the global flux can 
then be evaluated from the line EW. Assuming log~$g$\,=\,8, the measured EW of the \Hg line in
%Table~\ref{EWFW} 
Table 3 yields a dilution factor 
of (0.17$\pm$0.04) at $\sim$4500\,\AA\, for the range of temperatures determined above.
Given the steep energy distribution of the white dwarf, its contribution 
is significantly lower around \Ha with $\alpha \sim$0.06, corresponding to an expected 
absorption component of EW$_{abs}$ $\sim$ 1\,\AA. This is negligible compared to the emission 
component EW$_{em}$ $\sim$ 18\,\AA, therefore explaining the absence of noticeable 
absorption at \Ha.

The comparison of the observed WD flux with  
the predicted theoretical flux yields then an additional constraint 
on the R$_{wd}$/D ratio shown in Figure~4, where R$_{wd}$ is the white dwarf
radius in units of 10$^9$cm and D is the system distance in units of 100 pc.
For log~$g$\,=\,8.0, R$_{wd}$ is $\sim$0.87 10$^{9}$cm for a standard M/R relation (Nauenberg 1972)
and, from the range of temperatures (T = 18\,000-30\,000\,K) consistent with the observed FWHM,
the system distance is then derived as D = (210-250) pc.

\begin{figure}
\resizebox{\hsize}{!}
	{\includegraphics[angle=-90,bb=50 50 554 574,clip]{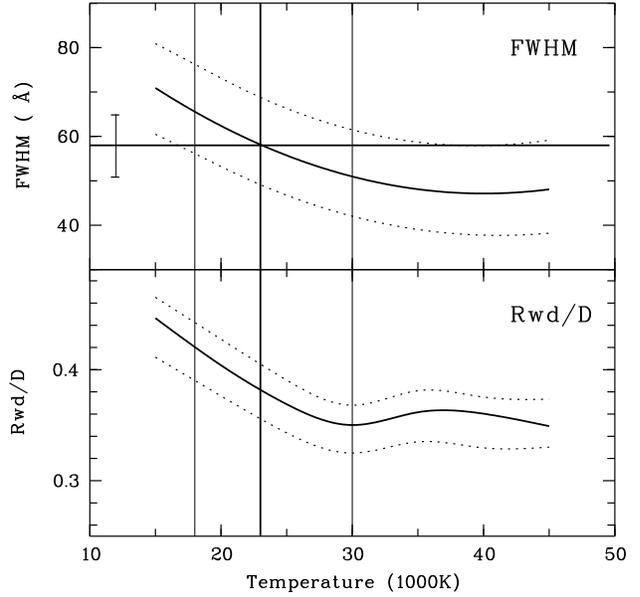}}
%\resizebox{\hsize}{!}{\includegraphics[angle=-90]{rx28fig4.ps}}
\caption{
Upper panel : The theoretical \Hg\, FWHM values predicted from white dwarf atmospheric
models for a range of temperatures
and different log~$g$ (log~$g$=7.5 upper dotted, 8.0 middle thick, 8.5 lower dotted).
The range of values consistent with the measured FWHM absorptions (51-65\,\AA)
and log~$g$=8.0 are shown by vertical lines. 
Lower panel : the R$_{wd}$/D values, 
in units of 10$^9$ cm for R$_{wd}$ and 100 pc for D,
as derived from a comparison of the observed and theoretical flux (see text)
for the same parameter range.
}
\label{fig4rel}
\end{figure}

The assumption of an underlying WD spectrum responsible for the observed 
broad absorption lines leads therefore to
a T$\sim$\,23\,000\,K white dwarf contributing $\sim$17\% (around 4500\,\AA) 
to the overall flux for a system at $\sim$230pc. 
These values are quite consistent with what can be expected from a typical IP system.
The absence of absorption associated with the He lines suggests
a pure DA white dwarf, at the difference of a disk emission where absorption helium lines
could be produced.
Another interesting feature is the absence of Zeeman splitting, such
as is observed in polars during low states. A significant Zeeman splitting is
expected and observed in isolated white dwarfs for magnetic fields greater than 6-10 MG
(see Achilleos et al. 1991) and is commonly present in polar systems with B fields in the range
10-30 MG. 
Such components are not detected here for \Hb or H$_{\gamma}$.
With the same method as described in Bonnet-Bidaud et al. (2000), synthetic
profiles have been built using the Zeeman wavelengths and oscillator strengths 
tabulated by Kemic (1974). The absorption features expected at the position of
the Zeeman components from a magnetic field as low as 3 MG have been compared to the
observed spectra. We find in particular that for the lowest tabulated B
value of 3 MG, the  \Hb $\sigma$ components at  4832\,\AA\, and 4896\,\AA\, are clearly missing.
The observations are therefore the first direct indication that, 
at least in some intermediate polars, the white dwarfs have
lower magnetic fields than in polars. A low magnetic field for
\rx seems also suggested by the short (312 s) spin period of the system if one assumes that
the system is at equilibrium with a spin period corresponding to the Keplerian period
at the Alfven radius (see Norton et al. 1999).

The question arises as to why \rx is, so far, the only system among IPs where the white dwarf is seen
to contribute significantly to the overall flux so that the atmospheric absorptions 
become visible. From the deduced best characteristics, there is no indication that
the white dwarf is atypical. Its temperature (T$\sim$\,23\,000\,K) is similar, 
though at the upper edge of the values reported for magnetic CVs (Sion 1999). 
However, the gravity being not fully constrained; the contribution of the white dwarf could be 
indeed more important if the gravity is somewhat lower. This would imply a larger radius together
with a higher temperature, as seen from Figure 4. For log\,$g$\,=\,7.5,  the white dwarf 
contribution around 4500\,\AA\,
would be larger by a factor $\sim$ 3.5 when compared to log\,$g$\,=\,8.0. This would however   
also correspond to a rather low mass 
object ($\sim$0.34M$_{\sun}$) with a very peculiar high temperature (T $\sim$ 35\,000\,K),
the lower limit being at T $\geq$ 25\,000\,K from the lower bound of the FWHM values.

A sufficient contrast between the white dwarf 
and the overall flux could also be obtained if the contributions from the other
regions in the system are significantly lower when compared to other intermediate polars.
From the residual (total-white dwarf) flux at a distance of $\sim$230\,pc, an absolute
magnitude of M$_{v}$$\sim$7.6 is derived, after correction for a visual extinction of
A${_V}$$\sim$0.45 corresponding to the N$_{H}$ $\sim 8.7\,10^{20}$ at.cm$^{-2}$ 
column density observed in the X-rays (de Martino et al. 2001a, submitted). 
If attributed to a disk, and using the crude (M$_{v}$-$\dot{M}$) relation given in Smak (1989) 
for face-on blackbody disks, this absolute magnitude would correspond, 
for a $\sim$0.6M$_{\sun}$ (log\,$g$\,=\,8.0) white dwarf, 
to a  $\dot{M}$ in the range  $\sim (1-2)\,10^{16}$ g.s$^{-1}$,
depending on the exact extension of the disk.
This is consistent with the range of values $\sim (0.7-1.3)\,10^{16}$ g.s$^{-1}$ derived from the recent X-ray spectral analysis
(de Martino et al. 2001a, submitted). If, as suggested from an analysis of the X-ray pulsations 
(Norton et al. 1999), the disk is seen at a relatively low inclination angle, its luminosity would 
therefore not be abnormaly low.

However, the possibility of a larger inclination angle cannot yet be excluded. 
The large dispersion seen in the radial velocity curves
(see sect. 3.3) is strongly suggestive of the S-wave line distorsions introduced by a variable hot spot 
contribution and/or partial eclipses of the line regions found in high inclination systems
(see Warner 1995b). There is also some evidence from recent photometric observations of 
a possible partial disk eclipse (de Martino et al. 2001b, in preparation).
We therefore cannot exclude that the important contribution of the white dwarf 
to the overall flux could be due to 
a combination of a low gravity white dwarf with a somewhat inclined disk.

Additional observations are also clearly needed to see if the visibility of the white dwarf also 
depends on the overall accretion rate and luminosity of the source.

%\section{Conclusion}

\medskip
\it Acknowledgments. \rm We are very grateful to D. Koester and N. Dolez for having provided
us with detailed catalogues of theoretical white dwarf spectra.

\end{document}